\documentstyle[prl,twocolumn,aps]{revtex}
\begin{document}
\draft
\title{Diamagnetic response due to localization in chains of connected
mesoscopic
rings}
\author{S. Kettemann$^{\left( 1 \right)}$\cite{weiz}, K. B. Efetov$^{(1,2)}$}
\address{$\left( 1\right) $ {\it Max-Planck Institut f\"urFestk\"orperforschung
}%
Heisenbergstrasse1, D-70569 Stuttgart, Germany\\$\left( 2\right) $%
Landau Institute for Theoretical Physics, Moscow, Russia}
\date{\today }
\maketitle

\begin{abstract}
A dynamic response to a magnetic field of a chain of connected mesoscopic
rings is considered. We show that the low frequency behavior corresponds to
localization of the electrons along the chain and to diamagnetic dynamic
currents inside rings. The magnetization density due to these currents does
not vanish even in the limit of the infinitely long chain of strongly
connected rings showing that this is a macroscopic effect. Being of a
dynamic origin the currents can be destroyed by inelastic scattering but we
argue that the corresponding decay time can at low enough temperature be
large compared to all other time scales of the system.
\end{abstract}

The recent measurement of the magnetization of an array of disconnected
Cu-rings \cite{p0} constituted the first detection of a thermodynamic
mesoscopic effect and revealed its dependence on the statistical ensemble.
Considerable effort has been taken in order to build a theory of the
observed effect in the diffusive regime where the circumference of a ring $L$
is larger than the elastic mean free path $l.$ Averaging over disorder at
fixed Fermi energy gives an amplitude for the persistent current which is
exponentially small in $L/l$ \cite{k2}, and it was suggested that a larger
amplitude can be due to the fixed number of particles in the rings \cite{r9}%
. However, averaging over impurities in the canonical ensemble can
analytically only be done by a perturbation theory in fluctuations of the
chemical potential \cite{r9}.

Alternatively, it was proposed that the averaged dynamic response to the
magnetic field is less sensitive to
the statistical ensemble \cite{ref13}.
Levels in a closed ring as functions of the external magnetic field do not
intersect each other and, before the averaging, the dynamic response must
coincide in the limit $\omega \rightarrow 0,$ where $\omega $ is frequency,
with the corresponding thermodynamic derivative. Therefore, in this case,
one can hope that calculating the dynamic response with a fixed chemical
potential at $T=0 K$, as was done in \cite{ref13}, one obtains a reasonable
description of the thermodynamic currents. Both the methods give the same
order of magnitude of the persistent current in the diffusive regime \cite
{ref13} and the same result in the quasi-ballistic regime\cite{p2}.
Although, possibly, not being exact, the dynamic approach can serve as a
reasonable approximation scheme for calculation of thermodynamic quantities.

However, let us emphasize that calculation of the dynamic response in the
isolated rings, due to the level repulsion, does not bring any new physics
with respect to thermodynamics and can serve as a calculational tool only.
(Recently, it was argued \cite{p3} that in a different experimental setup
the dynamic response can give results completely different from the
thermodynamic ones).

Now, while level intersections are
 not possible in an isolated ring, they can
occur in a chain of connected rings in the limit of $%
L_x\rightarrow \infty ,$ where $L_x$ is the length of the chain. This is due
to the fact that all wave functions in a disordered one dimensional chain
are localized and parts of the chain located far from each other do not
interfere. This leads to an exponential suppression of the level repulsion.
The resulting level intersections in the system cause the dynamic response
now to be completely different from the thermodynamic one. In fact, in the
infinite system one cannot expect any finite thermodynamic magnetization
density except the Landau diamagnetism which is rather small and will not be
considered here.

In this Letter we consider the dynamic response of a chain of connected
rings in the limit of infinite length $L_x$. We assume that every single
ring is in the diffusive regime. It will be shown that in this disordered
system the magnetization caused by an applied magnetic field is proportional
to the length $L_x$ of the chain and, depending on the tunneling amplitude
between the rings, can be measurably large.It decays in time due to inelastic
processes, only. This effect is intimately related to the phenomenon of
localization in quasi-one dimensional disordered systems. The system
considered below offers the unique possibility to study localization in a
magnetization measurement of an isolated system rather than in a transport
measurement of the system coupled to leads.

To understand better the nature of the dynamic response we want to study in
the diffusive regime, let us consider first the corresponding clean limit.
The chain of $N_r$ rings is in the ballistic regime $L<l$ an ideal diamagnet
and its magnetization due to an applied flux $\phi $
 through the rings is
given by:

\begin{equation}
\label{e1}M_{dyn}=N_rVK_d\phi _\omega /4\pi c^2,\qquad K_d=-ne^2/m,
\end{equation}
where $\phi _\omega $ is a small time dependent part of the flux through the
ring. In Eq. (1) $m$ is the electron mass, $e$ -the electron charge, $n$
-the electron density, $c$ - the light velocity and $V$ -the volume of a
single ring. This magnetization does not correspond to the free energy
minimum and due to inelastic processes decays to the smaller thermodynamic
value which can exceed the Landau diamagnetism, however \cite{arg}.

Let us now consider a model of non-interacting electrons moving in a random
potential. The impurity averaged response to an external magnetic field $%
\tilde H$
\begin{equation}
\label{e5}^{}\tilde H=H+H_\omega \cos (\omega t),
\end{equation}
is calculated using the Kubo formula in the form given in Ref. \cite{ref13}.
The frequency dependent magnetization $M_\omega $ is then obtained from the
response function $K(\omega )$ as
\begin{equation}
\label{mag}M_\omega =N_rV%
\mathop{\rm Re}
\left[ K(\omega )\phi _\omega \right] /4\pi c^2
\end{equation}
It will be shown below that in the absence of inelastic processes the
dissipative part of the magnetization vanishes in the zero frequency limit
but the real part of the response can remain finite. In the limit of small
frequency the diagrammatic technique leads to divergencies and one should
use the supersymmetry method \cite{susy}.

The responses both along the chain of rings with cross-section $S=V/L$, as
well as in the azimuthal direction of the rings can be reduced to a
calculation of functional integrals over supermatrices $Q$. Here, we want to
calculate the response in azimuthal direction of an ensemble of chains of
rings whose length exceeds the localization length $L_x>>L_c>>L$. The
dynamic response for the problem of persistent currents in disconnected
mesoscopic rings was calculated in Ref. \cite{ref13} using the zero
dimensional version of the $\sigma $-model\cite{ref13}. In the general case
of an arbitrary hopping amplitude between the rings one has to add in the
free energy functional a term describing the coupling between the rings.
Such a term can be written in the form analogous to the Josephson coupling
in the theory of superconductivity\cite{m1,m3}. Changing the probability of
tunneling from ring to ring we can describe thus the crossover from the case
of isolated rings to the homogeneously weakly disordered chain of rings.
Then the free energy $F\left[ Q\right] $ is written in the form
\begin{equation}
\label{bethe}F[Q]=-\sum_{ij}J_{ij}%
\mathop{\rm STr}
\left[ Q_iQ_j\right] +\sum_iF_0\left[ Q_i\right] ,
\end{equation}
\begin{equation}
\label{bethe1}F_0\left[ Q_i\right] =\frac{\pi \nu V}8\sum_i%
\mathop{\rm Str}
\left[ -{\cal D}_0\left( \frac ec{\ }{\bf A}[Q_i,\tau _3]\right) ^2+2i\omega
\Lambda Q_i\right] ,
\end{equation}
where
 $i$ and $j$ enumerate the rings in
the chain, $D_0$ is the classical diffusion coefficient and we remind that $%
Q_i^2=1$. The notation $%
\mathop{\rm STr}
$ stands for the Supertrace. The definition of the matrices $\tau _3$ and $%
\Lambda $ as well as details of the supersymmetry method can be found in
previous works \cite{susy,ref13}. The function $F_0\left[ Q_i\right] $ in
Eqs.(\ref{bethe},\ref{bethe1}) stands for electron motion inside the rings,
while the first term in Eq.(\ref{bethe}) describes the coupling between the
rings due to electron hopping. The parameter $J$ is related to the single
electron hopping amplitude $t$ as $J=\pi ^2V^2\nu ^2t^2.$ Only nearest
neighbors are coupled so that $J_{ij}=J$ for the neighbors, and $J_{ij}=0$
otherwise. The limit $J=0$ corresponds to the chain of disconnected rings.
In the limit $J\gg 1$ the model on the lattice Eq. (\ref{bethe}) becomes the
continuous one dimensional $\sigma $-model \cite{susy}. Then, the coupling
constant $J$ is related to the classical diffusion constant $D_0$ as
\begin{equation}
\label{e20}J=\pi \nu VD_0/8L^2.
\end{equation}

Due to the one- dimensionality of the model Eq.(\ref{bethe}) one can use the
transfer matrix technique. Corresponding partial differential equations have
been written in Refs.\cite{susy,k4} for the continuous $1D$ $\sigma $-
model. For the model on the lattice, Eq.(\ref{bethe}), the corresponding
recurrence equation is an integral one \cite{m1,m3}. Repeating the main
steps of these works we reduce calculation of integrals over $Q_i$ for all
sites $i$ to one integral over $Q$
\begin{equation}
\label{e21}\left\langle \ldots \right\rangle _Q=\int \left( \ldots \right)
\Psi ^2\left( Q\right) \exp \left( -F_0\left( Q\right) \right) dQ.
\end{equation}
The function $\Psi \left( Q\right) $ in Eq.(\ref{e21}) satisfies the
equation
\begin{equation}
\label{e22}\Psi \left( Q\right) =\int \exp (2JSTrQQ^{^{\prime }})\exp \left(
-F_0\left( Q^{^{\prime }}\right) \right) \Psi \left( Q^{^{\prime }}\right)
dQ^{^{\prime }}.
\end{equation}
The solution of the Eq.(\ref{e22}) depends on the vector potential $A$
entering $F_0$, Eq.(\ref{bethe1}). In principle, it can be solved, at least
numerically, for arbitrary magnetic fields using the parameterization
proposed recently \cite{iida}, but we will present results only in the
limits of zero and high enough magnetic fields. These limits correspond to
the orthogonal and unitary ensembles. In both cases one can omit the first
term in Eq.(\ref{bethe1}) because in the unitary case the supermatrix $Q$
commutes with the matrix $\tau _3$ since the time reversal symmetry is broken%
\cite{susy}. Due to the symmetry of the free energy the function $\Psi
\left( Q\right) $ depends only on one compact real integration variable $%
\lambda $ and two non compact integration variables $\lambda _{1,2}$
parameterizing the supermatrix $Q$ \cite{susy}. Therefore, in the functional
integral over $Q$ one can integrate first over all other variables of the
representation of $Q$ introduced in Ref. \cite{susy}, reducing it thus to an
integral over $\lambda $ and $\lambda _{1,2}$.

Calculations for arbitrary frequency are most simple for the unitary
ensemble, since the time reversal symmetry is broken and there is only one
non compact integration variable $\lambda _1$. Corresponding calculations
are not very different from those performed in Refs.\cite{ref13} and we
obtain
\begin{equation}
\label{e23}K^{unit}\left( \omega \right) =-i\omega \sigma _0\{1
\end{equation}
$$
+\frac 12\int_{-1}^1\int_1^\infty \exp \left[ \left( i\pi \omega /\Delta
-\delta \right) \left( \lambda _1-\lambda \right) \right] \Psi ^2\left(
\lambda ,\lambda _1\right) d\lambda d\lambda _1\}
$$
where $K^{unit}(\omega )$ is the response along the circumference of the
rings in the unitary ensemble, $\Delta $ $=\left( \nu V\right) ^{-1}$ is the
mean level spacing in one ring in the chain, $\sigma _0=e^2\tau n/m$ is the
Drude conductivity and $\delta \rightarrow 0$. There remains a finite
response after ensemble averaging, if at low frequencies $\omega $ the
second term in the brackets in Eq.(\ref{e23}) is proportional to $1/\omega $%
. In the limit $\omega \rightarrow 0$ solving Eq.(\ref{e22}) becomes more
simple because the main contribution to the integral over $\lambda _1$ comes
from $\lambda _1\sim 1/\omega $. Then, the function $\Psi \left( Q\right) $
depends only on one variable $z=2\omega \lambda _1$ and we obtain for the
response $K\left( \omega \right) $
\begin{equation}
\label{e24}K^{unit}\left( \omega \rightarrow 0\right) =-\sigma _0\Delta
_{eff}\left( J\right) /\pi ,
\end{equation}
where the effective mean level spacing $\Delta _{eff}$ is a non- trivial
function of the coupling $J$ between the rings
\begin{equation}
\label{e25}P\left( J\right) =\Delta _{eff}\left( J\right) /\Delta
=\int_0^\infty \exp \left( -z\right) \Psi _J^2\left( z\right) dz
\end{equation}
This function gives the probability that an electron does not leave a ring
forever and is known numerically for arbitrary $J$ \cite{m3}. Here, we have
to remark that the order in which the limits were taken, first taking the
length of the chain $L_x$ to infinity and then performing the zero-frequency
limit means that our results apply for chains of finite length only if the
frequency $\omega $ is not smaller than the mean level spacing of the chain $%
\Delta _0=1/\nu SL_x$.

At high frequencies the second term in Eq. (\ref{e23}) is small and one
obtains the classical response function $K(\omega )=-i\omega \sigma _0$. In
this limit the conduction is classical in all directions. We see from Eqs. (
\ref{e23}, \ref{e24}) that the characteristic frequency of the crossover
from the quantum to the classical regime is of the order of $\Delta _{eff}$.

In the limit $J=0$ corresponding to disconnected rings, $\Psi =1$ and $%
\Delta _{eff}\left( 0\right) =\Delta $. The magnetization of the chain of
disconnected rings is then, using Eq. (\ref{mag}):
\begin{equation}
M_\omega ^{unit}(J=0)=N_rVK_d\tau \Delta \phi _\omega /4\pi ^2c^2,
\end{equation}
We remind that for the isolated rings the possibility of the averaging with
the fixed chemical potential is a certain assumption while for $J\neq 0$ one
should consider the whole chain and in the limit $L_x\rightarrow \infty $
the difference between the canonical and grand canonical ensembles vanishes
(however, taking into account charging effects one can have effectively
isolated rings at finite $J).$

In the opposite limit $J\gg 1$ the function $\Psi \left( z\right) $ is the
solution of the differential equation
\begin{equation}
\label{e27}zd^2\Psi /dz^2-16J\Psi =0.
\end{equation}
Solving Eq.(\ref{e27}) and calculating the integral Eq. (\ref{e25}) we find
that there is still a finite probability $P(J)=1/(96J)$ that an electron
stays in one ring. This is a direct consequence of the localization of the
electrons along the chain. In the unitary ensemble the localization length $%
L_c$ was calculated in Ref.\cite{susy} and can be related with the help of
Eq.(\ref{e20}) to the coupling $J$ as $L_c=32JL$ where we included an
additional factor $2$ since the effective cross-section of the chain of
rings is $2S$. Note, that with an effective length of a ring $L/2$ in the
chain the inverse participation ratio becomes $\Gamma =2/3L_c$ which is the
well known result in the unitary ensemble. In this limit we obtain for the
effective level spacing $\Delta _{eff}\left( J\right) $
\begin{equation}
\label{e29}\Delta _{eff}\left( J\gg 1\right) =\left( 3\nu SL_c\right) ^{-1}.
\end{equation}
We see from Eq. (\ref{e29}) that in the limit of large $J$ the response $%
K^{unit}\left( 0\right) $ Eq. (\ref{e24}) looks as if the chain of $N_R$
rings consisted effectively of $L_x/3L_c(J)$ rings. In this limit we obtain
with Eq. (\ref{mag}) that the magnetization does not depend on the disorder:

\begin{equation}
\label{e30}M_\omega ^{unit}(J\gg 1)=\left( N_rV/(8\pi M_T)^2\right) K_d\phi
_\omega /4\pi c^2,
\end{equation}
where $M_T=k_F^2S/4\pi ^2$ is the number of transverse channels in a single
ring.

Decreasing the coupling $J$ makes the localization length shorter which
leads to a larger response.

The response $K$ Eq. (\ref{e24}) describes the dynamic magnetization per
unit length along the chain and therefore the total magnetization is
proportional to the length of the chain. It has been mentioned that the
corresponding thermodynamic response is small as well as the difference
between the canonical and grand canonical ensembles. To make a more explicit
estimate, one can use the expansion in terms of fluctuations of the chemical
potential \cite{r9,alt} and calculate the correlation functions entering the
expansion with the supersymmetry technique. It turns out that the
thermodynamic magnetization density
of the chains calculated in this approximation
 vanishes for
macroscopically long chains\cite{phd}.

In order to estimate the decay time of the dynamic magnetization one has to
specify inelastic processes. As we found out above,
 the dynamic magnetization is due to the localization along the chain.
As soon as the localization is lifted and electrons can travel
through the chain, the magnetization should vanish.
At temperatures $T > \Delta_{eff}$ and
at frequencies $\omega $
smaller than $t^{-1}\left( J\right) ,$ where $t\left( J\right) $ is a time
due to inelastic processes, the dynamic magnetization decays due to a
 d.c. conductivity proportional to $%
t^{-1}\left( J\right) $ whereas at higher frequencies $t(J)$ has no influence
on this magnetization.
 The simplest way to include the time $t\left( J\right) $ into
formulae obtained above is to substitute $\omega \rightarrow \omega
+it^{-1}\left( J\right) $ in all expressions for $K/\omega ,$ where $K$ can
stand for the responses in both the longitudinal and azimuthal directions. ($%
K/\omega $ is proportional to the product of two Green functions and it is
this quantity that is calculated in the linear response theory). With this
substitution the magnetization would decay with time $t\left( J\right) $. Of
course, this is a hypothesis and we suggest it to make a rough
estimate only. As concerns the longitudinal conductivity in the regime of
the localization it can be different from zero due to hopping between the
localization centers \cite{mott}. Explicit calculations for disordered
chains \cite{gmr} at  not too low temperatures
corresponding to $T > \Delta_{eff}$ lead to the result
$\label{md}t\left( J\right) =\tau _{ph}$
where $\tau _{ph}$ is the electron-phonon scattering time.  To the best of our
knowledge nobody has performed analogous calculations for thick wires which
are equivalent to the chain of the rings but we hope that this estimate
 can give a reasonable estimate also for the case considered here.
Then, the inverse decay time is proportional to $T^3$.
Thus, at temperatures much below the Debye temperature $\omega_D$ but above
the effective level spacing, $ \omega_D >> T > \Delta_{eff}$,
one can have $1/t(J) < \Delta_{eff}$ and it should be possible to
observe the large diamagnetic response.

At  lower temperatures $T\ll \Delta _{eff}$  the exponential
Mott law, $t(J) \sim (1/\omega_D)
\exp( 4 (3 \Delta_{eff}/T)^{1/2})$, might serve as a lower
limit for the decay time. Then, it increases
exponentially with decreasing the localization length.

We note that charging effects might increase both the
amplitude and the decay time of the dynamic magnetization.
A detailed analysis of these effects will be the subject of further research.

Analogous calculations can be
performed also for the orthogonal ensemble corresponding to the zero static
component $H$ of the magnetic field Eq. (\ref{e5}). In this case we obtain
as in Ref.\cite{ref13}
\begin{equation}
\label{e31}M_{\omega \rightarrow 0}^{orth}\left( 0\right) =0.
\end{equation}
Changing the static component $H$ of the magnetic field we can have a
crossover between Eqs.(\ref{e31}) and (\ref{e24}). The characteristic flux $%
\phi _c$ of this crossover is of the order of $\phi _0(\Delta
_{eff}/E_c)^{1/2}$, where $\phi_0= 2\pi c/e$ and
$E_c=\pi ^2D_0/L^2$ is the Thouless energy, $\phi
_0$ is the flux quantum. The dependence of the response on the flux $\phi $
is periodic with the period $\phi _0/2$.

Adding magnetic or spin-orbit impurities changes Eqs. (\ref{e24}, \ref{e31}%
). The system with the magnetic impurities corresponds to the unitary
ensemble. One can use as before Eq. (\ref{e24}) provided $\Delta
_{eff}\left( J\right) $ is substituted by $\Delta _{eff}\left( 2J\right) /2.$
If the magnetic impurities are absent spin-orbit ones lead to a different
function $\tilde \Delta _{eff}\left( J\right) $. However, this difference is
only numerical and does not change the sign of the response which is in all
cases diamagnetic.

In conclusion, we showed that a magnetic field applied perpendicular to a
chain of disordered connected rings induces a macroscopic diamagnetic
current. In the absence of inelastic scattering which can correspond to low
temperatures this current can live for a very long time. The corresponding
magnetization density remains finite even in the limit of a macroscopically
long chain of perfectly connected rings. At the same time the longitudinal
response along the chains shows a dielectric behavior usual for disordered
wires. In fact, the shorter the localization length along the chain is, the
larger is the current in the rings {\ and the larger is the relaxation time
to its vanishing thermodynamic value}. We suggest to check our theoretical
results by measuring the magnetization as a function of flux of chains of
mesoscopic rings in the diffusive regime or, alternatively, of linear
antidot lattices where the mean free path due to elastic scattering is
smaller than the distance between neighboring antidots. This can yield a new
contactless method of studying the localization.

Both of us are grateful to V. Falko for critical reading the manuscript and
useful advices. S. K. thanks the Weizmann Institute, where this work was
completed, for hospitality and Alex Kamenev, Nathan Argaman, Peter Kopietz,
E. Altshuler, Yoseph Imry, Alexander M. Finkel'stein, Yuval Gefen and Daniel
Braun for most useful discussions. S. K. acknowledges support from the
Minerva foundation.

\end{document}